\pdfoutput=1
\documentclass{iopart}
\usepackage{iopams,graphicx}  

\newcommand{\malpha}{\mbox{\boldmath{$\alpha$}}}

\begin{document}

\title[Diffusive transport without detailed balance in bacteria]{Diffusive transport without detailed balance in motile bacteria:\\
Does microbiology need statistical physics?}


\author{M. E. Cates}

\address{SUPA, School of Physics and Astronomy, University of Edinburgh, JCMB Kings Buildings, Mayfield Road, Edinburgh EH9 3JZ, UK}

\begin{abstract}
Microbiology is the science of microbes, particularly bacteria. Many bacteria are motile: they are capable of self-propulsion. Among these, a significant class execute so-called run-and-tumble motion: they follow a fairly straight path for a certain distance, then
abruptly change direction before repeating the process. This dynamics has something in common with Brownian motion (it is diffusive at large scales), and also something in contrast. Specifically, motility parameters such as the run speed and tumble rate depend on the local environment and hence can vary in space. When they do so, even if a steady state is reached, this is not generally invariant under time-reversal: the principle of detailed balance, which restores the microscopic time-reversal symmetry of systems in thermal equilibrium, is mesoscopically absent in motile bacteria. This lack of detailed balance (allowed by the flux of chemical energy that drives motility) creates pitfalls for the unwary modeller. Here I review some statistical-mechanical models for bacterial motility, presenting them as a paradigm for exploring diffusion without detailed balance. I also discuss the extent to which statistical physics is useful in understanding real or potential microbiological experiments. 
\end{abstract}

\maketitle

\section{Introduction}
Bacteria are unicellular organisms, capable of self-reproduction and, in many cases, motility (the biologists' term for self-propulsion). A range of biomechanical mechanisms for this motility are shown by different species of bacteria, or in some cases by the same species under different environmental conditions. 

Among the simplest of these mechanisms is the swimming motion of species such as {\em Escherichia coli} (the most studied bacterium of all) \cite{bergbook}. Individual {\em E.~coli} have helical flagella -- whiplike appendages -- each of which is forced to rotate by a biochemically powered motor located where it meets the cell body. (The cell body of an {\em E.~coli} is about 2$\mu$m long and 0.5 $\mu$m wide; its flagella about 10 $\mu$m long and 20 nm wide.) Because of the chirality of these flagella, their clockwise and anticlockwise motion is inequivalent. One sense of rotation causes the flagella to form a coherent bundle which acts like a (low Reynolds number analogue of a) ship's propeller, resulting in a smooth swimming motion. Initiating the other sense of rotation -- even in just a subset of the bundled flagella -- causes the bundle to separate, with the result that the cell starts to rotate randomly. The canonical motion of {\em E.~coli} then consists of periods of straight line swimming (called `runs') interrupted by brief bursts of rotational motion (called `tumbles'), see Figure 1. Tumbles are controlled by a biochemical circuit that throws flagellar motors into reverse gear every so often. A typical run lasts about 1s; to a reasonable approximation the duration of runs is Poisson-distributed, so that tumbles can be viewed as random events occurring with a certain event rate $\alpha$. Tumbles are usually much shorter, of duration $\tau\simeq 0.1$s and often treated as instantaneous ($\tau \to 0$). In practice, tumbles may not totally randomize the orientation of the cell body, but for simplicity that assumption is often made, and we adopt it here unless otherwise stated.

\begin{figure}
\begin{center}
\includegraphics[width=55mm]{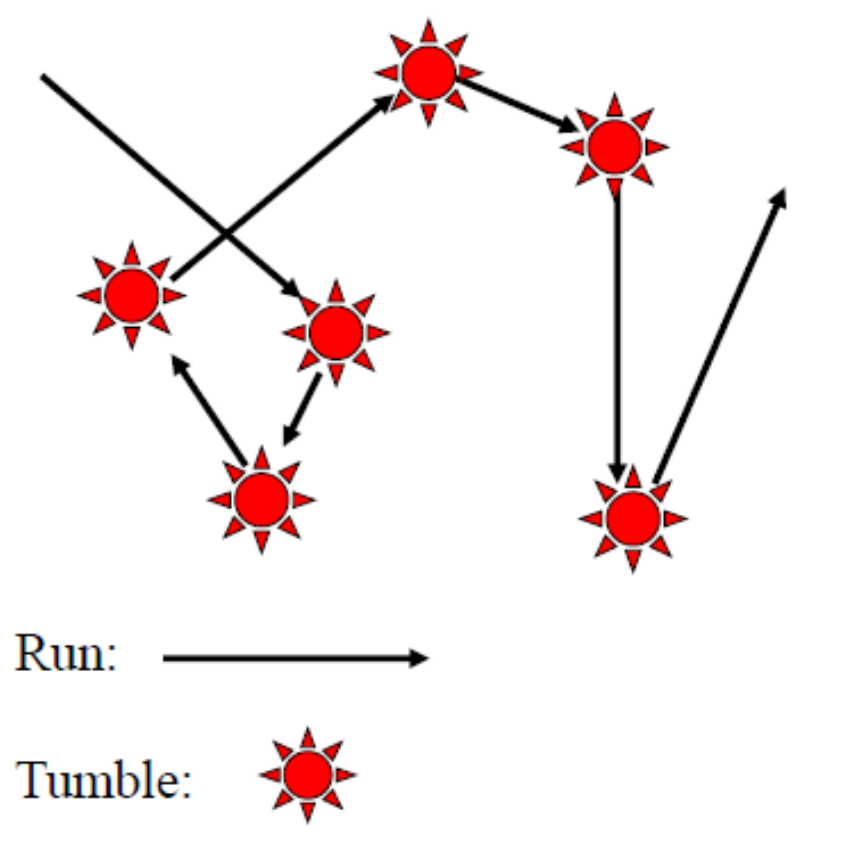}
\caption{Schematized run-and-tumble dynamics of {\em E.~coli}.}
\end{center}
\end{figure}

The swim speed $v$ of {\em E.~coli} (and other similar species such as {\em Salmonella typhimurium}) is is around 20$\mu$m/s, so that under the simplest conditions -- in which the statistics of the run-and-tumble motion is unbiased by any environmental factors -- each bacterium performs a random walk, with an average step length of about 20$\mu$m, and about one step per second. The diffusion constant of this stochastic process is readily calculated, and (setting $\tau\to 0$ for simplicity) obeys
\begin{equation}
D = \frac{v^2}{\alpha d} \label{D}
\end{equation} 
where $d$ is the spatial dimensionality. This diffusivity is hundreds of times larger than would arise by the Brownian motion of colloidal particles of the same size -- as can be experimentally checked using a deflagellated mutant (see \cite{Jana}). On the other hand, Brownian motion does still matter: it sets the longest duration possible for a straight run before its direction is randomized by rotational diffusion. The run duration chosen by evolution is comparable to, but shorter than, the rotational diffusion time $\tau_B$, so that runs are fairly straight. 

At time- and length-scales much larger than $1/\alpha$ and $v/\alpha$ respectively, the dynamics just described is equivalent to an unbiased random walk, which is in turn equivalent to force-free Brownian motion with constant diffusivity $D$. The run-and-tumble motion of bacteria requires metabolic energy (a food source) and hence represents a far-from-equilibrium process entirely dependent on internal fluxes.
Nonetheless, in the absence of environmental factors causing parameters such as $v$ and $\alpha$ to vary in space, at mesoscopic length and time scales the process maps onto free Brownian diffusion, which exhibits detailed balance. The same does not apply when the motility parameters do vary in space. Such variation creates a biased diffusion process which, in the general case, does not have detailed balance. This contrasts with Brownian motion of colloids in an external potential (such as gravity) or with conservative interparticle forces: in those cases, detailed balance is still present and thermal equilibrium (the Boltzmann distribution) is achieved in steady sate.
In thermal equilibrium there can be no circulating fluxes of any kind, whereas for motile bacteria it is easy to construct steady-state counterexamples to that rule: we will encounter several below. For this reason, bacterial motility is offered in this article as an interesting paradigm for diffusion without detailed balance, while already comprising an experimentally well-studied and scientifically important topic in its own right.

For microbiological purposes, the most important example of spatially varying run-and-tumble parameters is probably chemotaxis \cite{chemobook}. This is the main mechanism whereby bacteria navigate their environment. As described in more detail below, to achieve chemotaxis bacteria use a biochemical circuit that, roughly speaking, computes the change $\Delta c$, on a fixed time scale $\tau_c$, in the local concentration $c$ of a chemoattractant such as food. If this is positive, meaning that the organism is swimming in a good direction, then the tumble rate $\alpha$ is decreased from its free-swimming value $\alpha_0$. Put differently (and anthropomorphically!), the organism dynamically creates an estimate of ${\bf v}.\nabla c \simeq \Delta c/\tau_c$, and uses this vectorial information to decide how long to keep moving in the direction of ${\bf v}$ \cite{segall,block}. 
To maximize efficiency, $\tau_c$ and $1/\alpha$ must both be as large as possible. However, were the run time $1/\alpha$ to exceed the Brownian rotation time $\tau_B$, or were $\tau_c$ to exceed the run time, the swimmer's attempt to translate temporal into spatial information -- which depends on runs being straight -- would fail. Thus one expects $\tau_c \simeq 1/\alpha_0 \simeq \tau_B$, and evolution has indeed arranged things this way. 
As well as navigating towards better environments (and away from less favourable ones) bacteria may use chemotaxis to detect each others' presence, by sensing a molecule emitted by other individuals. Although the strategic origins of the chemotactic response mechanism have recently been explored from a statistical physics viewpoint \cite{strong,PGG,clark,kafri}, most of the literature on the macroscopic consequences of chemotaxis postulates or develops models at a more coarse-grained level, as we will mainly do from now on.

Chemotaxis is not the only way bacterial motility can be altered by environmental factors; some of the alternatives are explored below. For instance many bacteria modify their behaviour directly in response to the local concentration of signalling molecules (as opposed to their gradients), including those emitted by other individuals. An example is the `quorum sensing' response which causes changes of phenotype (such as a transition to a virulent state) once the local concentration of bacteria exceeds a pre-set threshold \cite{quorum,biofilms}.

One other important way in which bacteria respond locally to levels of food or signalling molecules is by cell division (when food is plentiful)
or cell death (when it runs out). The resulting dynamics is often modelled by a logistic-type equation for $\rho$, the local bacterial density, as
\begin{equation}
\dot\rho = A\rho(1-\rho/\rho_0)\label{logistic}
\end{equation}
where the `target density' $\rho_0$ depends on the environmental conditions, and represents the highest population level sustainable under those conditions. If the density is less than $\rho_0$ the population grows, if it is above $\rho_0$ it decays. In practice cells might not die, but merely cease to breed and/or enter a non-motile dormant state; we ignore such complications here. 

There is a distinction between studying the biochemical aspects of microbiology (addressing control circuits, signalling pathways etc.), and studying microbes from the viewpoint of collective behaviour, diffusion, pattern formation, and hydrodynamics. The latter domain is our primary concern in what follows. Within that domain, we can broadly distinguish between two approaches. One is taken traditionally by mathematical biologists, and generally involves setting up deterministic differential equations at population level. Such equations are often targeted at a quantitative description of specific datasets; they sometimes, though not always, involve numerous fitting parameters.
The power of this approach, across vast swathes of biology (from embryology through cancer growth to ecology), is surveyed by Murray \cite{murray}.
 A complementary approach, which is the main focus here, is grounded in statistical physics \cite{newman}. It emphasises the role of stochasticity; the identification of phase transitions in parameter space; and the use of minimal models to explore universal, or at least generic, mechanisms -- even when this seriously compromises a model's ability to quantitatively fit the data. The relevance or otherwise of this approach to real microbiology is discussed at the end of the article.  

\section{Run-and-tumble models for independent particles}\label{independent}
We start in 1D with an idealized model as formulated by Schnitzer and others \cite{schnitzerberg,schnitzer}. As befits a physics-oriented approach, we solve this model first in idealized situations before, in later sections, exploring its relevance to some (real or potential) microbiological experiments. The idealized situations include sedimentation equilibrium (particles subject to a uniform external force) and trapping in a harmonic well. 

Consider a single particle confined to the $x$ axis, and let $R(x,t)$ and $L(x,t)$ be the probability densities for finding it at $x$ and moving rightward or leftward respectively. Allow both the swim-speeds $v_{L,R}$ and tumble rates $\alpha_{L,R}$ to be different (in general) for left- and right-moving particles, and assume tumbles to be of negligible duration. Note that in 1D, half of all tumbles are ineffective in changing the direction of motion (they convert $R$ into $R$ or $L$ into $L$). Then, since tumbles are independent random switching events one has (with overdot denoting $\partial/\partial t$ and prime $\partial/\partial x$) 
\begin{eqnarray}
\dot R &=& -(v_RR)'-\alpha_RR/2 +\alpha_LL/2 \label{Rdot}
\\
\dot L &=& (v_LL)'+\alpha_RR/2 -\alpha_LL/2 \label{Ldot}
\end{eqnarray}
These equations can be exactly solved in steady state for any specified dependences of the four parameters $v_{L,R}$ and $\alpha_{L,R}$ on the spatial coordinate $x$.

They can also be systematically coarse-grained to give a diffusion-drift equation for the one-particle probability density $p \equiv R+L$:
\begin{equation}
\dot p = (Dp'-Vp)' \label{pdot}
\end{equation}
where explicit forms relating $D(x)$ and $V(x)$ to $v_{L,R}(x)$ and $\alpha_{L,R}(x)$ are given in \cite{JTPRL}. These forms are chosen to ensure that all steady states of (\ref{Rdot},\ref{Ldot}) and of (\ref{pdot}) are identical. (Transient behaviour will differ of course, since at short times there is less information in (\ref{pdot}) than in (\ref{Rdot},\ref{Ldot}).) 

The resulting steady states have some notable features. For instance, in the symmetric case, where $v_L = v_R = v(x)$ and $\alpha_L = \alpha_R = \alpha(x)$, one finds the steady-state density \cite{schnitzer}
\begin{equation}
p_{ss}(x) = p_{ss}(0)\frac{v(0)}{v(x)} \label{inverse} 
\end{equation}
where the origin $x=0$ has been chosen as an arbitrary reference point. Thus, the probability density for symmetric run-and-tumble particles is inversely proportional to their speed, but independent of their tumble rate. (The latter holds for instantaneous tumbles only; at finite tumble duration $\tau$, increasing $\alpha$ is equivalent to decreasing $v$ \cite{schnitzer,JTEPL}.)

To those statistical physicists whose intuition has been developed mainly in the context of equilibrium systems, the $v$-dependence in (\ref{inverse}) is quite strange. There is no force on these particles, so they have no potential energy. The Boltzmann distribution for isothermal Brownian particles, even with a spatially varying diffusivity $D(x)$, would have $\rho_{ss}$ independent of $x$, in contradiction to (\ref{inverse}). 
(Spatially varying diffusivity arises, for instance, when colloids move in a medium of nonuniform viscosity.) 
On the other hand, to shoppers on the high street the result (\ref{inverse}) is intuitive: like pedestrians, bacteria are unconstrained by detailed balance, and accumulate wherever they move slowly (for instance, in front of an interesting shop window).  

A second intriguing result concerns sedimentation. In a sedimenting system, upwards and downwards swimmers have different speeds (obeying in 1D $v_{L,R} = v\pm v_s$, where $v_s$ is a sedimentation velocity). The exact result in this case is 
\begin{equation}
p_{ss}(x) = p_{ss}(0)\exp[-x/\lambda] \label{sed}
\end{equation}
where the decay length obeys $\lambda = (v^2-v_s^2)/\alpha v_s$ \cite{JTEPL}. The exponential form is  the same as found by Perrin for Brownian colloids under sedimentation \cite{Perrin}; this is no different from the isothermal atmosphere (a gas under gravity) whose exponential density profile features in most undergraduate physics courses. However, in both of those equilibrium cases, the decay length is inverse in the strength of gravity, $\lambda= D/v_s$ (with $v_s = Dmg/k_BT$, where $m$ is the buoyant mass), so the thickness of the sedimented layer smoothly goes to zero as gravity goes to infinity. In contrast, for independent run-and-tumble particles, the layer thickness goes to zero as $v_s\to v$: complete collapse occurs at a finite threshold of gravity. This result, like the previous one, is unsurprising on reflection: for $v_s>v$, even
the upward-swimming particles are moving downwards, so that there can be no steady state unless all particles are in contact with the bottom wall. The same calculation can be done in 3D, with the same conclusion (but a different functional form for  $\lambda$) \cite{JTEPL}. 

Some readers may complain that this result depends on our having a fixed propulsion speed $v$, while in practice of course there is some distribution of speeds \cite{wilson}. Were this to be a Maxwell-Boltzmann distribution (with each bacterium somehow sampling this ergodically) then a truly Boltzmann-like result might be recovered. But that requires the distribution of swim speeds to extend, albeit with very small probability, to unbounded values. As far as we know, this is not physiologically possible: bacteria, unlike colloids or gas molecules, have some nonrelativistic maximum swim speed $v_m$, and collapse of the one-particle probability density must then still occur at $v_s> v_m$. Residual Brownian motion will change this in principle, but the resulting $\lambda$ is then almost the same as for immotile microbes (and typically of order one particle diameter or less).

Similar results arise when a run-and-tumble particle is confined to a harmonic potential.
Such a particle cannot escape beyond an `event horizon' at $r^*$ -- the radius at which its propulsive speed, when oriented outward, is balanced by the speed at which it is being pulled inward by the confining force. For large tumble rates or weak trapping ( $\alpha r^*\gg v$) the particle rarely ventures out towards $r^*$ and a near-gaussian steady state density at the trap centre is attained. (The system then equates to a Brownian particle of matching diffusivity $D$, confined in the same potential.)
In contrast, when tumbling is rare, a particle starting anywhere within the trap will soon arrive at $r^*$
and (essentially) wait there until its next tumble. In consequence, the steady state probability density has its maximum no longer at the centre of the trap, but at $r^*$ \cite{JTEPL}. (Such a state, in 3D, is visible as the opening frame of Figure 5 below.) 

In both the above examples, where the swim speed of a single run-and-tumble particle varies with its position and/or swimming direction, the behaviour predicted in 1D and higher dimensions are qualitatively similar (though not identical). However a strong, qualitative dependence on dimensionality arises when not the swim speed but the tumble rate $\alpha$ depends on position and swimming direction (represented in 3D by the unit vector ${\bf \hat v}$). Schnitzer \cite{schnitzer} assumed a low-order multipole expansion such that $\alpha = \alpha_0 +\malpha_1({\bf r}).{\bf\hat v}$. He found that so long as the vector field $\malpha_1$ is conservative ($\malpha_1 = \nabla \phi$), as always holds in 1D but does not hold generally in higher dimensions, then the scalar from which it derives serves as an effective potential in a Boltzmann-like description:
\begin{equation}
p_{ss}({\bf r}) \propto\exp[-\phi({\bf r})/v] \label{alphass}
\end{equation}
On the other hand, if $\nabla\times\malpha_1 \neq 0$, as can arise for $d>1$, then the steady state contains circulating currents and no mapping onto a thermal equilibrium system is possible. 

\section{Chemotaxis} 
An important application of the above ideas, and the main motivation for studying them in much of the modelling literature \cite{schnitzer,othmer1,othmer2,rivero}, is chemotaxis. As outlined previously, chemotactic bacteria can move themselves up a chemical gradient $\nabla c$, where $c$ is the concentration of a chemoattractant, by modulating their tumbling activity. For a chemorepellant, the sign of the effect is reversed and the arguments given below must be modified accordingly. 

The standard description is to work directly at a coarse-grained level, adopting the diffusion drift equation for the one-particle probability density (\ref{pdot}), and phenomenologically asserting that the drift velocity $V$ in that equation obeys $V=\chi\nabla c$, with $\chi$ some constant (that may depend on $D$). A theory that relates $\chi$ to the microscopic run-and-tumble parameters has had to await relatively recent advances in our understanding of how chemotaxis works microscopically \cite{segall,block,PGG}. 

A simplified picture of this understanding is as follows. A bacterium has an on-board biochemical circuit whose job is to modulate the tumble rate according to the following integral
\begin{equation}
\alpha = \alpha_0-\int_{-\infty}^t K(t-t')c(t')dt' \label{int}
\end{equation}
Here $\alpha_0$ is value arising when $\nabla c = 0$. 
This baseline value can depend on environmental factors, but is broadly independent of overall shifts in the concentration of chemophore $c$ \cite{block,leibler}. This implies that $\int K(t)dt=0$. Moreover, $K(t)$ is a bi-lobed function (Figure 2) which essentially computes the change in local concentration $\Delta c$ experienced by the bacterium over a certain time scale $\tau_c$. If this change is positive, the next tumble is delayed proportionately.
For simplicity we have assumed in (\ref{int}) that a negative change conversely promotes tumbles, although experiments in fact suggest a delay-only, one-sided response. (Correcting this would lead to factor 2 changes in some of the results below.) The characteristic time $\tau_c$ sets the temporal scale for the bi-lobed $K(t)$ response (Figure 2). If this time were much longer than $1/\alpha_0$ under normal conditions (in most species, ``normal" means bacteria swimming in water), then the time integral in (\ref{int}) would cease to be informative of whether the swimmer is pointing up or down the chemical gradient. If on the other hand $\tau_c$ were extremely short, then detection of weak concentration changes would become unnecessarily noisy and inefficient \cite{bergbook}. Accordingly, one expects evolution to have set $\tau_c\alpha_0\sim 1$, and this is indeed the case \cite{block}.

\begin{figure}
\begin{center}
\includegraphics[width=75mm]{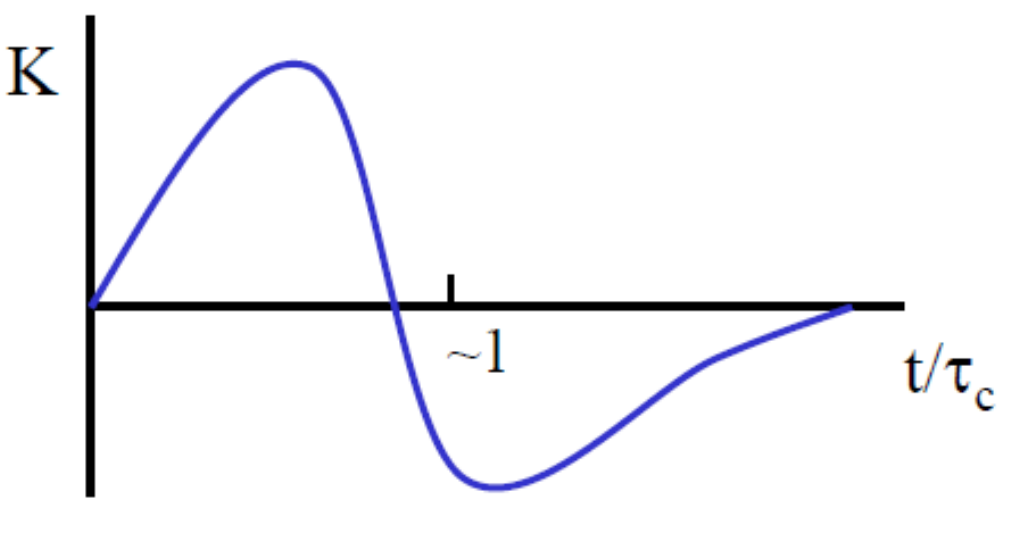}
\caption{Schematic depiction of the kernel $K(t)$ controlling the chemotactic response of {\em E.~coli}.}
\end{center}
\end{figure}

Expanding (\ref{int}) in weak concentration gradients, and assuming straight runs, one finds from the above considerations that 
$\alpha(t) = \alpha_0-\beta{\bf v}.\nabla c$, where (for a more careful analysis see \cite{PGG,Otti})
\begin{equation}
\beta = -\int_0^\infty \exp[-\alpha_0t]K(t)dt \label{beta}
\end{equation}
The exponential term inside the integral uses the unperturbed tumble rate to approximate the probability that the swimmer has not yet tumbled; once it does so, there is no longer a correlation between the temporal change of $c$ and its spatial gradient. 
The result implies $\malpha_1 = -\beta v\nabla c$ and hence $\phi/v = -\beta c$. It follows that steady-state chemotaxis (in which $c$ is by some external means maintained constant in time, but nonuniform in space) can be mapped onto a Boltzmann equilibrium problem via (\ref{alphass}), with $-c$ playing the role of a potential and $\beta$, defined via (\ref{beta}), playing the role of inverse temperature: 
\begin{equation}
p_{ss}({\bf r}) \propto \exp[\beta c({\bf r})]\label{chemoss}
\end{equation} 
To match this result for the probability density using the classical approach (comprising Eq.(\ref{pdot}) with $V = \chi\nabla c$) we clearly must choose $\chi = \beta D$.  

Constancy of $\beta$ in space is sufficient to ensure the absence of steady state currents, but these are present whenever $\nabla\beta \times \nabla c\neq 0$. On the other hand, for bacteria with identical phenotypes and hence identical $K(t)$, the $\beta$ value found from (\ref{beta}) can still vary between experiments performed in different media (through environmental influences on $\alpha_0$), even if it is spatially uniform in each experiment. This  observation will prove crucial in Section \ref{Chemo2}. 

It should be noted here that there are significant subtleties associated with chemotaxis which are not necessarily captured by
the adoption of spatially varying tumble rates in Eqs.(\ref{Rdot},\ref{Ldot}). Indeed, the biochemical mechanism outlined above cannot literally translate into spatially varying tumble rates for left- and right- moving particles since the time integral in Eq.(\ref{int}) continues to be calculated; half the time a tumble leads to no change of direction, so the future dynamics of a particle does depend on its history prior to the last tumble event. (The same applies, {\em mutatis mutandis}, in higher dimensions.) As lucidly discussed in \cite{Kafri11}, this becomes particularly important when determining exactly how the choice of kernel $K(t)$ influences chemotactic efficiency, both dynamically and in steady state \cite{strong,PGG,clark,kafri,Kafri11,vergassola}. This question is not our focus here, and will not be considered further below.

\section{Rectification}
Before moving on to address many-body physics, we consider one further intriguing property of bacterial motion that appears even at the level of a single particle. This is the phenomenon of rectification \cite{Austin1,Austin2,Austin3}. Working with bacteria in a two-dimensional layer, Austin and collaborators introduced a perforated wall made of asymmetric barriers (Figure 3). Partly for hydrodynamic reasons (see Section \ref{hydro}), a swimming bacterium that encounters a straight wall at an oblique angle tends (at least in this quasi-2D geometry) to continue swimming along the wall until either the wall ends, or the swimmer experiences its next tumble event. (Let's call this the `wall hugging' tendency.) Therefore a barrier as depicted in Figure 3 will act asymmetrically: it will funnel bacteria that approach from one side through the barrier, while causing those approaching from the other side to bounce off. Thus, if such a `funnel barrier' divides a container in two, the bacterial motion is on average rectified, causing an unequal steady state particle density on the two sides of the wall \cite{Austin1,Austin2}.

\begin{figure}
\begin{center}
\includegraphics[width=110mm]{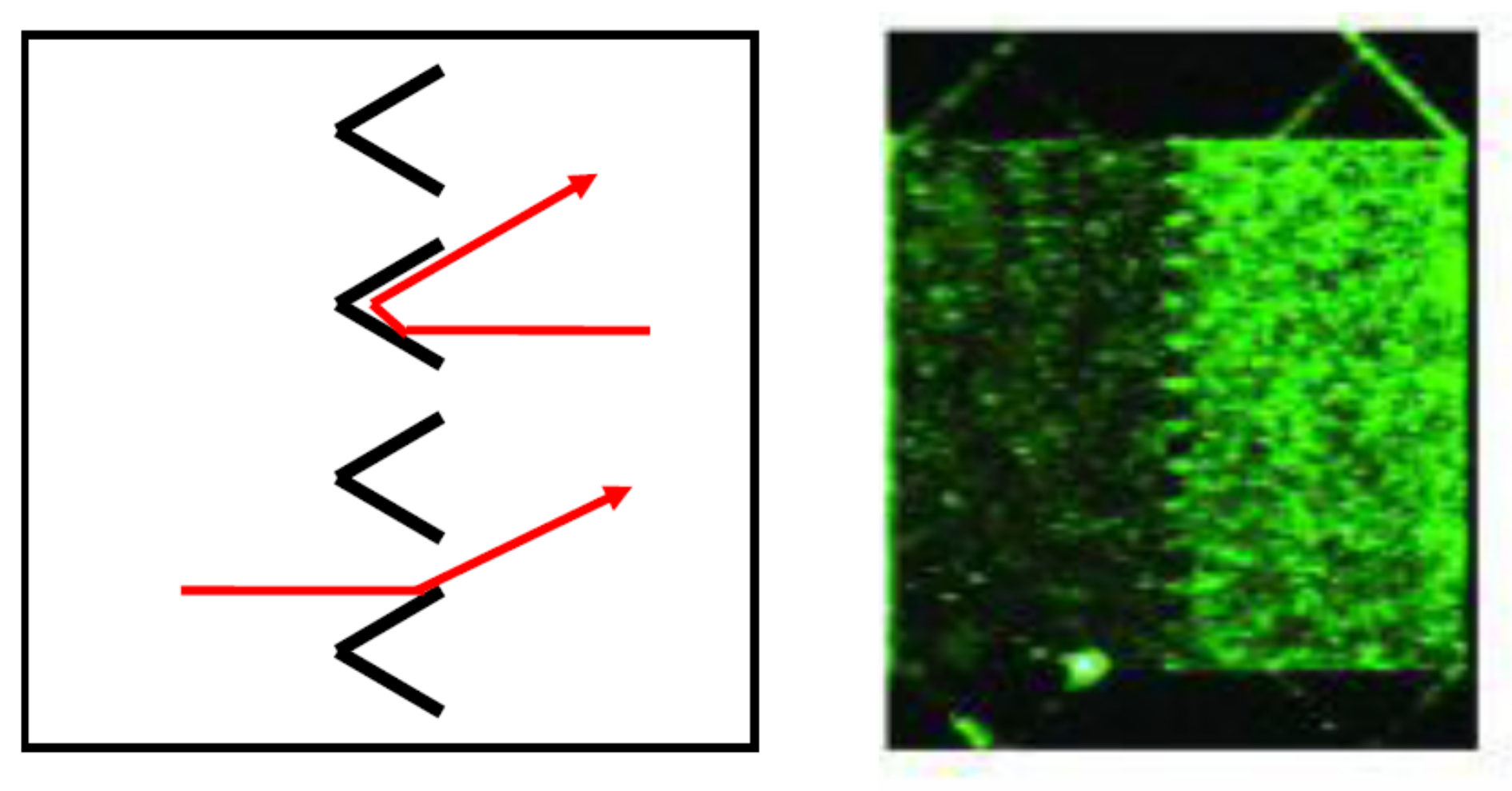}
\caption{Left: schematic effect of a funnel barrier. Note that the time reversal of either trajectory is highly improbable (it requires the particle to accidentally be aligned with the wall while still distant from it) breaking time reversibility. Right: fluorescent bacteria in a rectifying chamber, showing a steady state inequality in number density (from \cite{Austin1}).}
\end{center}
\end{figure}

A well known rectification theorem \cite{rect} states that such asymmetry can only emerge if spatial symmetry breaking (provided by the wall) is accompanied by time-reversal asymmetry in the trajectories. The latter is provided by the wall-hugging tendency (Figure 3) \cite{wan,JTEPL}. The necessity of such a mesoscopic violation of detailed balance can be confirmed by replacing the wall-hugging rule by an elastic collision law in direct simulations of the run-and-tumble dynamics; when this is done, rectification disappears \cite{JTEPL}. (This is true for both specular and `bounce-back' collisions \cite{wan2}.)

The simplest model capturing the rectification effect is to replace the wall by a strip on which the tumble rates for left- and right-moving particles are different. (This is reasonable, since particles approaching the wall from the repelling side end up getting turned back, which is like having an extra tumble, whereas those coming in the other direction can glide through with relatively minor angular deflection.) This model can then be addressed using the analysis of spatially varying $\alpha$ outlined in Section \ref{independent} above. For a single wall we then find from (\ref{alphass}) that the density ratio between the two sides of the wall obeys $p_1/p_2 = \exp(-\Delta\phi/v)$ where $\Delta\phi$ is, in a 1D version of the problem, simply the integrated difference in tumble rates as one passes across the strip \cite{JTEPL}. In three dimensions, the expression for $\Delta\phi$ is more complicated; nonetheless, the steady-state density ratio should depend only on this local property of the wall, not on the length of the wall nor on the shapes and sizes of the regions that it separates.

A corollary of this picture is that funnel gates could be used to create geometries where $\nabla \times \malpha_1$ is nonzero, so that the steady state contains macroscopic currents and no mapping onto a thermal equilibrium system exists (Figure 4). This was not explored experimentally in \cite{Austin1,Austin2} (though see \cite{Austin3}), but a directly related phenomenon, with exactly the same microscopic origin (the combination of a spatially asymmetric wall and time-irreversible swimming trajectories) has been reported subsequently. This comprises the unidirectional rotation of an asymmetrically saw-toothed rotor when immersed in a bath of bacterial swimmers \cite{rotorpaper1,rotorpaper2,rotorpaper3}, see Figure 4.  

A somewhat different interaction between swimming bacteria and funnel-like obstacles is reported in \cite{goldstein1}, wherein individual {\em B. subtilis} bacteria are shown to be reverse direction on encountering a narrow constriction without turning of the cell body (i.e., without tumbling as such). Unlike the much larger funnel gates used in \cite{Austin1,Austin2} these constrictions have dimensions of order the cell diameter. To achieve rectification, the rate for this flagella-flip process would have to be different for approaching a constriction from opposite sides; this seems perfectly possible for the case of an asymmetric, funnel-shaped obstacle.

\begin{figure}
\begin{center}
\includegraphics[width=110mm]{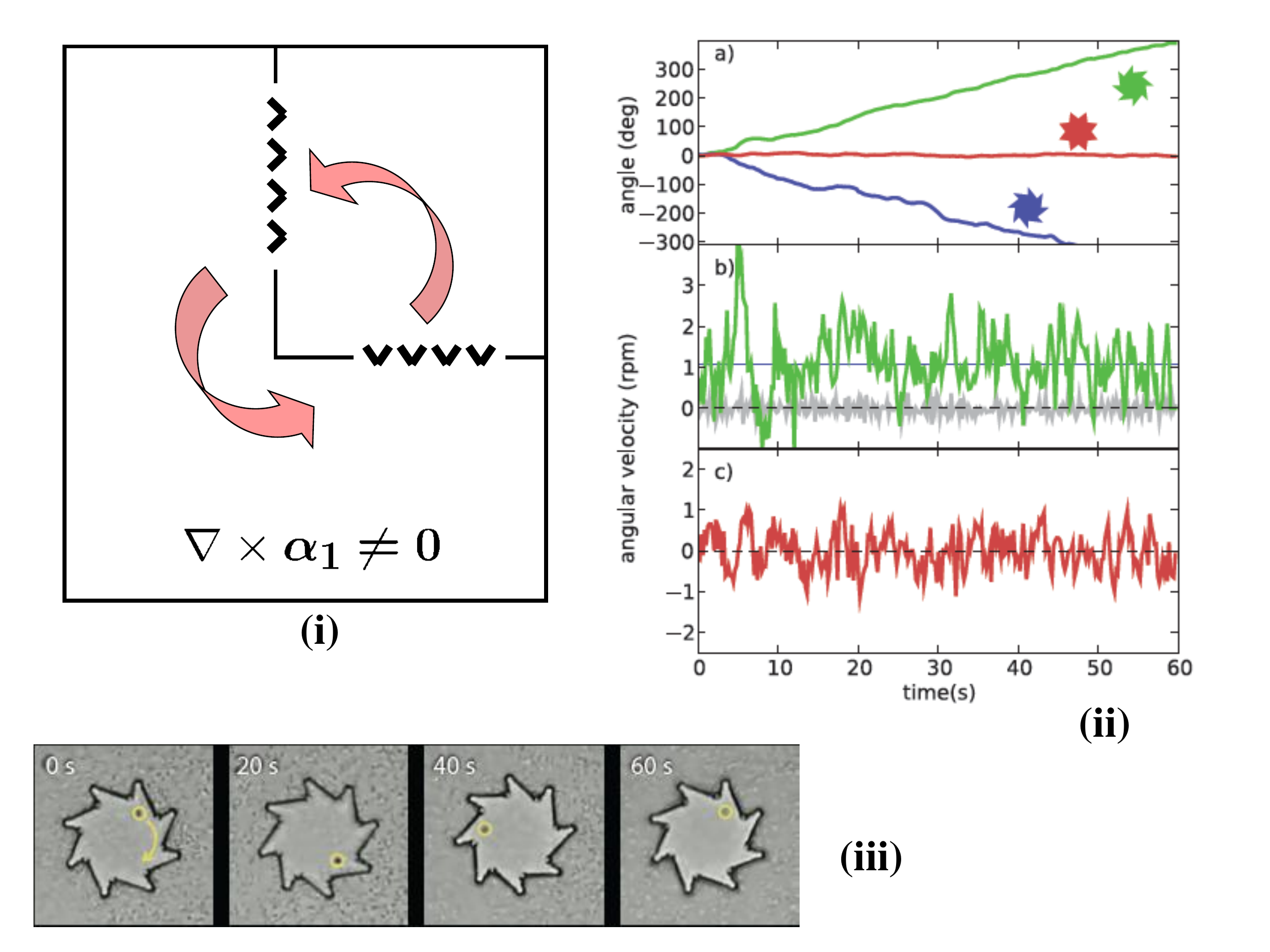}
\caption{(i) Funnel gate arrangement that would give a circulating current in steady state. (ii,iii) Experimentally observed angular rotation of an asymmetric cog in a bacterial bath (from \cite{rotorpaper1}).}
\end{center}
\end{figure}

\section{Hydrodynamic interactions}\label{hydro}
So far we have considered the motion of independent bacteria, focusing primarily on their steady-state probability density $p_{ss}$ under various environmental conditions. We now turn to interactions, starting with the hydrodynamic force that is exerted on one particle in proportion to the velocity of another swimming nearby. Such hydrodynamic interactions (HIs) are in general too complex to treat analytically; however we address them first because {\em hydrodynamic interactions have no consequences whatever for steady state densities in thermal equilibrium systems}. (Such densities obey the Boltzmann distribution, which is oblivious to any HIs that may be present.) Accordingly, any effect of HIs on the steady-state behaviour of bacteria is directly attributable to the lack of detailed balance in their mesoscale dynamics. 

Examples of this principle are already established at one-particle level where HIs can arise between a single bacterium and a wall; in conjunction with the chiral character of the flagellar propulsion system, this causes wall-bound {\em E.~coli} to have persistently spiralling trajectories \cite{chiral,loewen}. This chiral motion, which represents an average steady state current in violation of detailed balance, has been proposed to explain the observed tendency for bacteria to swim upstream against the flow of fluid down a pipe \cite{urethritis}, although a simpler, non-chiral mechanism, involving only the rotation of a swimmer's axis by the shear flow itself, appears equally plausible \cite{RupertPRL}. Such upstream swimming tendencies may help explain the invasive capabilities of pathogenic bacteria in infecting the urinary tract \cite{catheter}.

The swimming motions of bacteria also produce {\em interparticle} HIs, but these are relatively weak at large separations (and easily swamped by noise \cite{goldstein1a}). This is because swimming exerts a force dipole on the surrounding fluid, not a monopole as would be the case for a particle being dragged through the fluid by an external force. (Such monopole contributions are, of course, still present when swimming bacteria are additionally subjected to external forces such as gravity.) Hence the HI between swimmers falls off at large distances like $1/r^2$ instead of the $1/r$ result for motion induced by body forces \cite{lauga}. For bacteria, the force dipole is oriented to pull fluid in around the waist of the particle and eject it in both the forward and backward directions; combined with the orientational tendency of rodlike particles in shear flow, this `extensile' behaviour creates a negative contribution to the shear stress \cite{sriram}. Thus a dilute aqueous bacterial suspension can have a viscosity less than that of water \cite{dilute}; at high concentration the viscosity might in principle vanish altogether in laminar flow \cite{suzanne1,suzanne2}. The latter result is among many that were recently found from a collective hydrodynamic description of aligned bacterial fluids \cite{sriram,kruse} whose further description, with that of related experiments \cite{goldstein2,cisneros} and simulation studies \cite{shelley1,shelley2,pedley,wolgemuth} lie beyond the scope of this report.

Hydrodynamic interactions between swimmers also include important near-field terms, arising at separations comparable to the particle size. These depend in detail on the self-propulsion mechanism, and can lead to many intriguing phenomena ranging from relatively simple flock formation \cite{Llopis} to the phase-locking of nearby particles to form `synchronized swimming' teams \cite{synchswim}. (Near-field effects also control the swim speed $v$, which therefore need have no relation to the amplitude of the extensile force dipole.)
Perhaps unsurprisingly, this level of complexity offers serious challenges to computational researchers \cite{lauga,pedley,graham}. 
However, one avenue is to set up a minimal (far-field) numerical model of run-and-tumble swimmers in a continuous fluid medium, and use this to address the idealized physics problems discussed previously: sedimentation, and confinement in a harmonic trap. 

For sedimentation, one finds by this route that HIs have only limited effects \cite{RupertPRL}. The particle density $\rho_{ss}$ in steady state still decays exponentially with height (to numerical accuracy); however, there is some softening of the singularity associated with gravitational collapse when $v_s \to v$. This is understandable since any layer of collapsed particles will hydrodynamically set up a random stirring of the fluid that can allow at least some upward-pointing swimmers to briefly exceed the escape velocity. A much more drastic effect of HIs is found for particles confined in harmonic traps. Here the current-free steady state of inverted probability density (with the maximum at the outermost edge of the trap $r^*$ rather than the centre), as computed previously for the single-particle case at small $\alpha$, ceases to exist when the number of particles in the trap is high. This is because the shell of outward-swimming particles at $r^*$, now coupled together by hydrodynamics, is mechanically unstable to fluctuations in local density (Figure 5). The final result is collapse of the shell into a dense swarm, which resides at some distance $r_s<r^*$ from the centre of the trap. This swarm is almost stationary, and must therefore transmit the external force on all its members (provided by the trapping force) directly to the surrounding fluid. This fluid must accordingly be in motion: our system of self-propelled particles has spontaneously self-assembled into a pump \cite{RupertPRL}. More generally, it seems likely that HIs are more disruptive to a steady state in which swimmers are arranged in a coherent pattern (as is the case for the trap at small $\alpha$) than when they are locally disoriented and swimming in random directions (as holds at large $\alpha$, and in sedimentation).    

\begin{figure}
\begin{center}
\includegraphics[width=90mm]{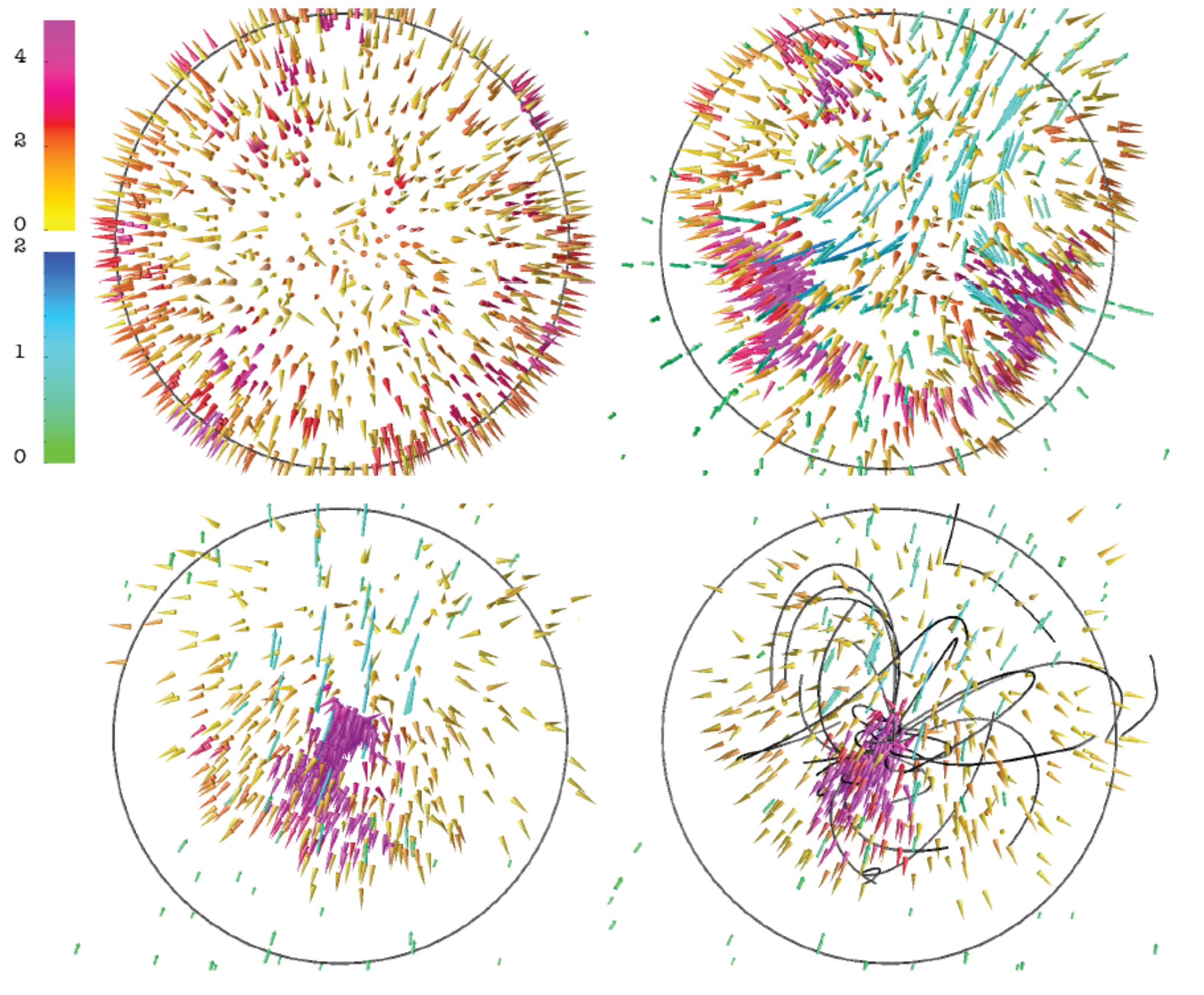}
\caption{Time series showing hydrodynamic instability of a shell of trapped swimmers in the low tumble rate regime. Swimmers are colour-coded yellow-to-magenta by local density; fluid velocity vectors are colour-coded green-to-blue by magnitude (scales top left, arbitrary units). The circle is the event horizon $r^*$ in the absence of hydrodynamics; the hydrodynamic simulation is initiated (top left) from a member of the steady-state ensemble in the absence of interactions, where the shell of particles at $r^*$ is visible. Black lines in the bottom right figure show trajectories of representative swimmers when a tumble causes them to leave the self-assembled, pump-like structure. Image courtesy of R. Nash; see \cite{RupertPRL} for a similar figure and full simulation details.}
\end{center}
\end{figure}

Before closing this discussion of many-body sedimentation and trapping, a brief comment is warranted on the feasibility of testing such predictions experimentally. In sedimentation, the interesting regime is $v_s\sim v$; this sedimentation velocity is higher than for bacteria in terrestrial gravity but the required range should easily be achieved by centrifugation. The literature is short of sedimentation studies, though one has been done for synthetic swimmers (whose dynamics is, however, not run-and-tumble) \cite{bocquet}. For particles in traps, the interesting regime is when the trapping radius $r^*$ is comparable to the run length $v/\alpha$. This requires a very soft trap compared to those generally achieved by optical methods, although suitable machinery may now be available \cite{cambridge}. From the viewpoint of fundamental physics it would be very worthwhile to see such predictions tested in detail -- while recognizing that they do not represent the kinds of question that most microbiologists would consider important.

\section{Density-dependent motility}
From now on we neglect hydrodynamic interactions and focus on systems in which run-and-tumble parameters such as $v$ and $\alpha$ vary in response to the local density of bacteria. Unlike all cases where these parameters are known in advance as a function of position (so that bacteria can be treated independently), this is a true many-body problem. Another class of problems, in which self-propelled bacteria also interact by direct colloidal interaction forces (such as the depletion interaction) is similarly interesting but not addressed here \cite{JTPRL,Jana2}. We also do not review in detail related studies of activity-induced phase separation in non-tumbling bacteria \cite{peruani1,peruani2} or active subcellular networks \cite{kruse00,marchetti}.

The first goal in addressing the many-body case is to derive an equation for the collective density field $\rho({\bf r}) = \sum_ig({\bf r}-{\bf r}_i)$ where the sum is over $N$ particles and $g$ is, in principle, a delta function. (For practical purposes we coarse-grain this spiky density by introducing a finite range to $g$.) 
A widespread procedure in the literature on noninteracting particles is to merely assert that $\rho = Np$ where $p$ is the one-particle probability density obeying (\ref{pdot}). This is incorrect, since $p$ is a probability density, not the actual density of one particle (which remains a delta-function); and while $p$ evolves deterministically, the collective density $\rho({\bf r})$ as defined above does not. One must therefore either define a probability density in the $N$-body configuration space and derive a Fokker-Planck equation at that level \cite{biroli}, or work with the Langevin 
equations (which are stochastic ordinary differential equations) for $N$ particles. The latter is more amenable to the handling of interactions: we can set up the Langevin equations with spatially varying run-and-tumble parameters and then allow this variation to occur through a functional dependence on the density field \cite{JTPRL}. This $\rho$-dependence passes smoothly from the $v,\alpha$ parameters to the one-body diffusivity $D([\rho],x)$ and drift velocity $V([\rho],x)$, which remain as previously defined in connection with (\ref{pdot}). These known functionals of density then enter a many-body, functional Langevin equation for the collective particle dynamics. Omitting an unimportant self-density term \cite{JTPRL} this reads in 1D:
\begin{equation}
\dot\rho = \left(-\rho V + D\rho' + (2D\rho)^{1/2}\Lambda\right)' \label{J_C} 
\end{equation}
where $\Lambda$ is a unit white noise. Clearly the final (noise) term goes missing if one simply asserts that $\rho = Np$ and then uses (\ref{pdot}) \cite{Dean,FrenchGuy}.

\begin{figure}
\begin{center}
\includegraphics[width=110mm]{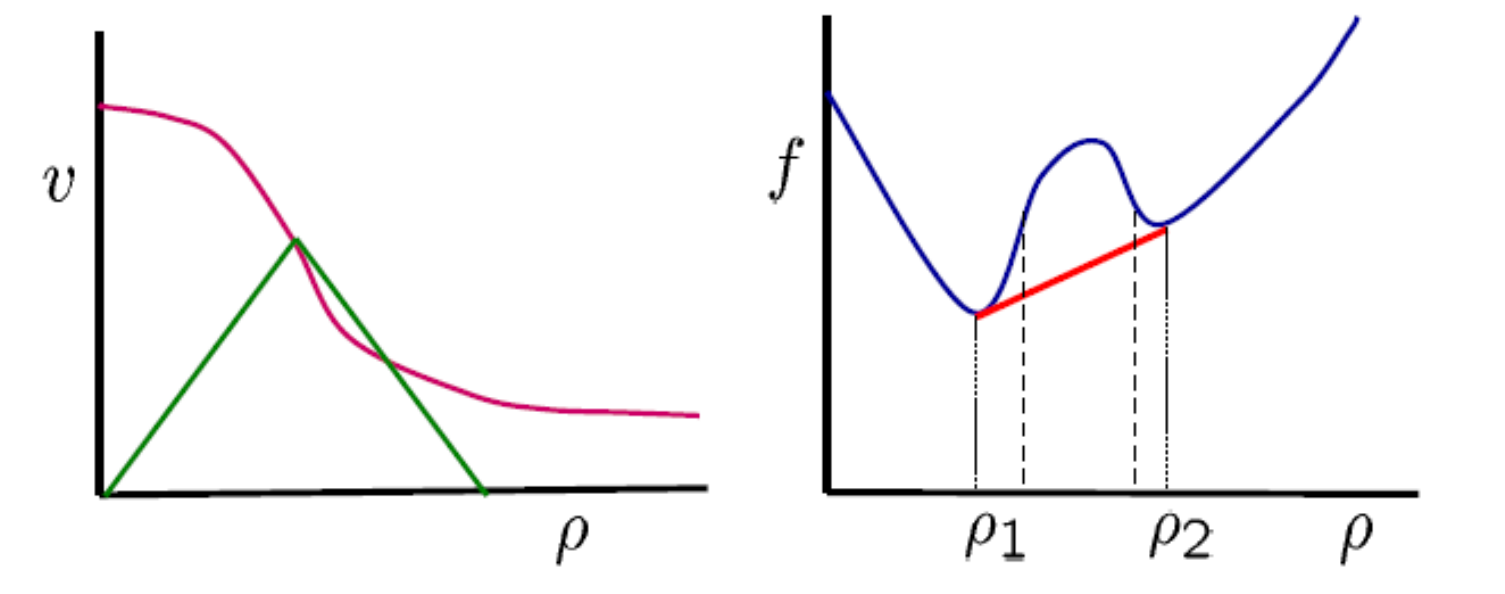}
\caption{Construction of the effective free energy density $f(\rho)$ in the mapping from 1D run-and-tumble particles with local motility interactions onto a fluid of interacting Brownian particles. If $v(\rho)$ decreases rapidly enough (left) the resulting $f(\rho)$ has a negative curvature (spinodal) region (right) with the global equilibrium state comprising a coexistence of the binodal densities $\rho_1,\rho_2$. The condition for instability ($f''<0$) translates into the geometric construction shown on $v(\rho)$: draw a line from the origin to any point on the curve and reflect this line in the vertical axis. If the slope of $v(\rho)$ is less than the reflected line, the system is unstable. See \cite{JTPRL}.}
\end{center}
\end{figure}

The upshot of this formal procedure is to establish that the interacting run-and-tumble system can be mapped at large scales onto a set of interacting Brownian particles, with detailed balance, if and only if a functional ${\cal F}_{ex}[\rho]$ exists such that
\begin{equation}
V([\rho],x)/D([\rho],x) = - [\delta {\cal F}_{ex}(\rho)/\delta\rho(x)]'
\label{functional}
\end{equation}
in which case, the system behaves like a fluid in which ${\cal F}_{ex}$ is the excess free energy. 
In general no such functional exists, but an exception is the case where $v(\rho)$ and $\alpha(\rho)$ are the same for both right- and left-moving particles, and depend on density in a purely local way. In that case one finds that the system is equivalent to a fluid whose free energy density (with $k_BT = 1$) is
\begin{equation}
f(\rho) = \rho(\ln\rho-1)+ \int_0^\rho \ln v(u)du \label{ff}
\end{equation}
On this basis we can conclude that when $v(\rho)$ is a sufficiently rapidly decreasing function of $\rho$, the local free energy density $f(\rho)$ of the equivalent thermodynamic system has negative curvature in an intermediate range of densities \cite{JTPRL} (Figure 6). At these densities, such a system is predicted to show a spinodal instability, separating into domains of two coexisting binodal densities $\rho_1$ and $\rho_2$, corresponding to a common tangent construction on $f$ (Figure 6).  In 1D these domains coarsen to a scale that, for systems with detailed balance, is finite and set by the interfacial tension, which in turn depends on gradient terms in the free energy functional. Such terms are neglected when one assumes a purely local dependence of $v,\alpha$ on density, and since the exact mapping onto a detailed-balance system is so far established only in that limit, they could allow coarsening to continue indefinitely even in 1D. Whether or not that applies, the initial spinodal instability and breakup into coexisting domains is an unmistakable physical effect, and is seen numerically both in 1D \cite{JTPRL} and 2D (Figure 7) \cite{Alasdair}. In these numerical experiments, domains also form by a different but equally well known mechanism (nucleation and growth) in the density range that lies between the binodal and the spinodal. (This lies between dashed and solid vertical lines in the plot of $f(\rho)$ presented in Figure 6.) In this range the system is stable to local perturbations but unstable globally; to capture the required nucleation events, the noise term in (\ref{J_C}) is essential.

\begin{figure}
\begin{center}
\includegraphics[width=65mm]{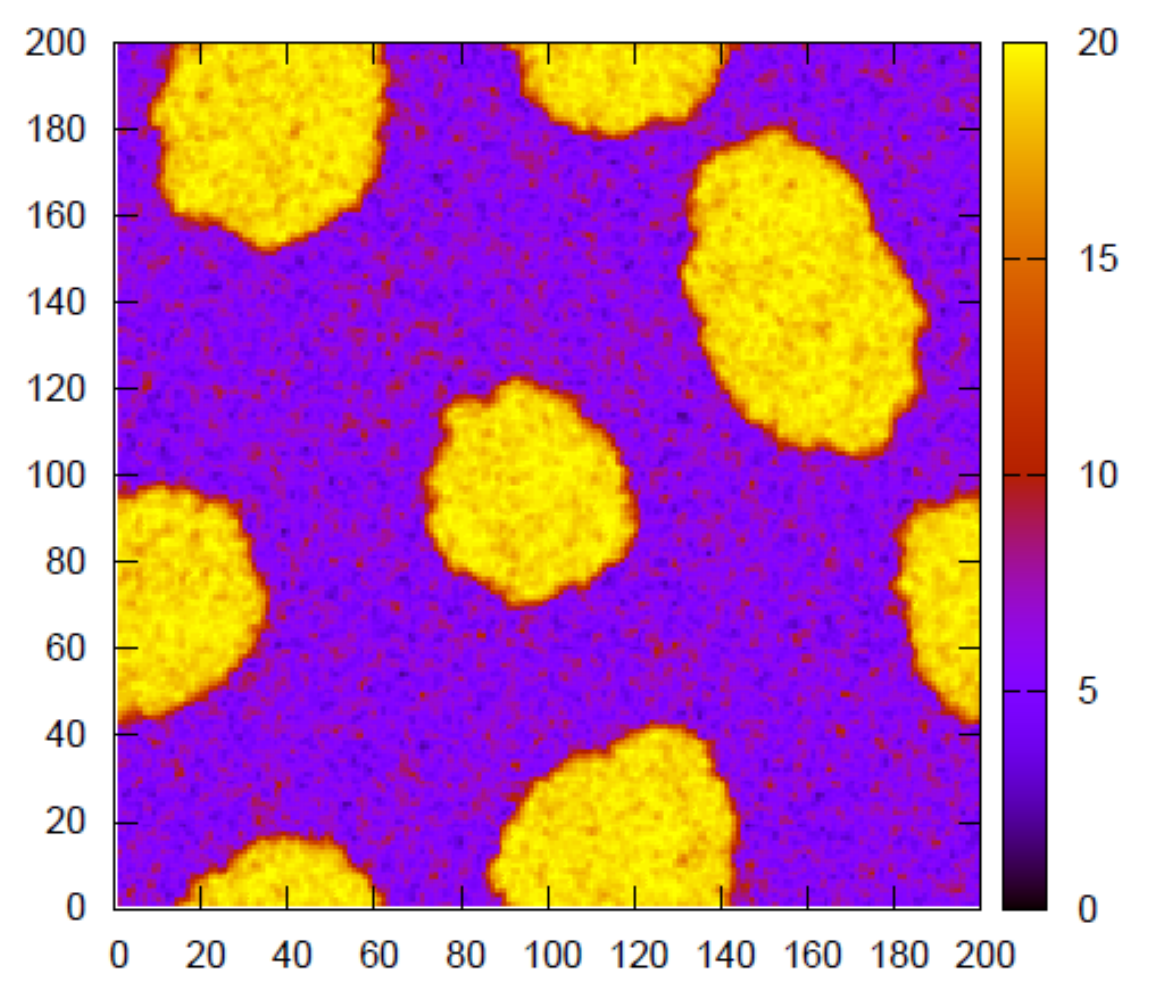}
\caption{A 2D run-and-tumble system undergoing motility-induced phase separation. Simulated via a lattice model ($200\times 200$ sites) as detailed in \cite{Alasdair}, with local density (particles per site) colour-coded on scale at right. Image courtesy of A. Thompson.}
\end{center}
\end{figure}

The physics behind such motility-driven phase separation is clear, and generic. We have seen earlier in Eq.(\ref{inverse}) that run-and-tumble particles accumulate (increase $\rho$) wherever they slow down (reduce $v$). But if $v$ is a decreasing function of density $\rho$, this means they also slow down on encountering a region of higher than average density. This creates a positive feedback loop which, so long as $v(\rho)$ decreases fast enough, runs away to phase separation. There is no attractive interaction between our diffusing run-and-tumble particles, yet they behave exactly as if there was one. 

In the microbiological literature, formation of dense clusters from a uniform initial population is often encountered (and usually called `aggregation' rather than phase separation). Certainly there are many situations where bacteria down-regulate their swimming activity at high density: for example this is fundamental to the formation, from planktonic swimmers in dilute suspension, of a biofilm \cite{biofilms}. A biofilm comprises a region with a high local density of bacteria that are immobilized on a wall or similar support. (In most biofilms, bacteria not only stop swimming but actually lose their flagella apparatus after a period of time.) Biofilms are ubiquitous, and generally unwanted; they arise in contexts ranging in seriousness from malodorous breath, via bacterial fouling of water supply pipes \cite{fouling}, to lethal infections in patients with cardiac valve implants \cite{valves}.

Biofilm formation generally involves chemical communication between individuals, but the effect of this may still be representable, at least in crude terms and during the initial stages, as a density-dependent swim speed $v(\rho)$. (A density dependent tumble rate $\alpha(\rho)$ or duration $\tau(\rho)$ has similar effects so long as the latter is finite.) Equivalent phase-separation physics could equally well occur by non-biochemical means, such as a simple crowding effect causing a reduced propulsive efficiency, hyrodynamically driven accumulation of bacteria near surfaces \cite{RupertPRL}, or by an intermediate mechanism such as secretion of a polysaccharide that increases the local fluid viscosity
(which is something that biofilms also can do \cite{exopoly}). 

\section{Population dynamics}
So far we have considered a range of similarities and differences between run-and-tumble motion and the Brownian diffusion of colloidal particles. 
However, bacteria have a further trick up their collective sleeve that colloids do not, namely self-replication. Alongside the transport of particles by a diffusion-drift process (described in general by the evolution equation (\ref{J_C})), the birth and death of bacteria leads to changes in particle density that are (unlike (\ref{J_C})), not the divergence of a current. A simple model of the population dynamics is the logistic equation given in Eq.(\ref{logistic}), and we stick to this model here while accepting its limitations. This equation describes a population that evolves, from above or below, towards a target density $\rho_0$ which for simplicity is presumed constant, and set by externally controlled environmental factors such as nutrient levels. 

This population dynamics can now be phenomenologically coupled to an equation for the phase separation dynamics in a system whose $v(\rho)$ decreases fast enough to make it unstable. (See \cite{Martin} for a similar example that instead involves sedimentation.) For simplicity we neglect dynamical noise (though keep it in the initial conditions) and assume that mildly nonlocal functionals $V([\rho],{\bf r})$ and $D([\rho],{\bf r})$ in (\ref{J_C}) (and in its higher dimensional analogues) generate a $\nabla^3\rho$ term in the current.  (This is the leading order allowed by symmetry.) The remaining local dependences $V(\rho)$ and $D(\rho)$ then together define a collective diffusion constant $D_c(\rho)$ (obeying $D_c(\rho)= \rho D \partial^2f/\partial\rho^2$ under the conditions where
(\ref{functional},\ref{ff})
apply)
 in terms of which we have \cite{PNAS}
\begin{equation}
\dot\rho = \nabla(D_c\nabla\rho) - \kappa \nabla^4\rho + A\rho(1-\rho/\rho_0)\label{PNAS}
\end{equation}
Throughout the range of spinodal instability, $D_c(\rho)$ is negative: accordingly small fluctuations in the initial density are amplified by the first term, and damped at short length scales by the second. On their own, as previously described, these terms would lead to phase separation into binodal phases of density $\rho_1$ and $\rho_2$. (The term in $\kappa$ effectively defines an interfacial tension for these domains and this influences their growth rate in the late stages, once interfaces become sharp.) However, this process of indefinite coarsening is clearly impossible, in any dimension, if the logistic term has a target density $\rho_0$ obeying $\rho_1<\rho_0<\rho_2$. This admits only one uniform steady state ($\rho = \rho_0$), so the system cannot evolve into large uniform patches with $\rho = \rho_1$ and $\rho = \rho_2$ as it would otherwise do. Instead, the outcome is a `microphase separation' whereby the spinodal pattern ceases to coarsen beyond a specific length scale. This length scale is fixed by the balance of the coarsening tendency against the fact that regions of low density are constantly producing new particles which must then diffuse into regions of high density, where they die off. A steady state domain pattern is then possible, which however does not map onto any equilibrium system as it contains mesoscopic particle currents from dilute to dense regions \cite{PNAS}.

The model just presented is rather general, and broadly agnostic as to the mechanism whereby $D_c$ has become negative in the spinodal region. All that is needed is a sufficient tendency for bacteria to move towards regions where they are already more numerous. As already explained, a sufficiently decreasing $v(\rho)$ can achieve this, but so can other mechanisms including conventional colloidal attractions \cite{Jana2}, or a quorum sensing response \cite{quorum}. Chemotaxis could also have the required effect, especially if the chemoattractant is produced by bacteria themselves, and has a short lifetime and/or high diffusivity. This combination of factors would create a nonlocal but near-instantaneous functional dependence of $V=\chi\nabla c$ on the bacterial density $\rho$, from which (\ref{PNAS}) then follows in the weak gradient limit. 

\section{Chemotactic patterns without chemotaxis}
The above remarks are pertinent to a classic set of microbiology experiments on multi-ring and spotted pattern formation in bacterial colonies innoculated from a point source (Figure 8) \cite{rings}. These patterns have long been taken by various workers as evidence for chemotaxis, and can indeed be reproduced by a detailed multi-parameter model in which bacteria explicitly secrete a chemoattractant in response to food (or another `stimulant', which itself diffuses) and then navigate up the gradient of the attractant \cite{murray,budrene,tyson}. One piece of evidence that these are indeed `chemotactic patterns' is that if the ability to secrete `chemoattractant' is disabled, the patterns go away \cite{confirm}. However, this finding does not rule out a chemically mediated but non-chemotactic response in which bacteria change their behaviour in response to the mere presence of the relevant chemical (quorum sensing \cite{quorum}) rather than swimming up its gradient (true chemotaxis). Moreover, whatever the actual mechanism in the organisms studied so far (
primarily {\em E.~coli} and {\em S.~tyhphimurium}) it is reasonable for a physicist to ask how general these patterns are, and whether their origin can be understood in simple mechanistic terms. 

\begin{figure}
\begin{center}
\includegraphics[width=85mm]{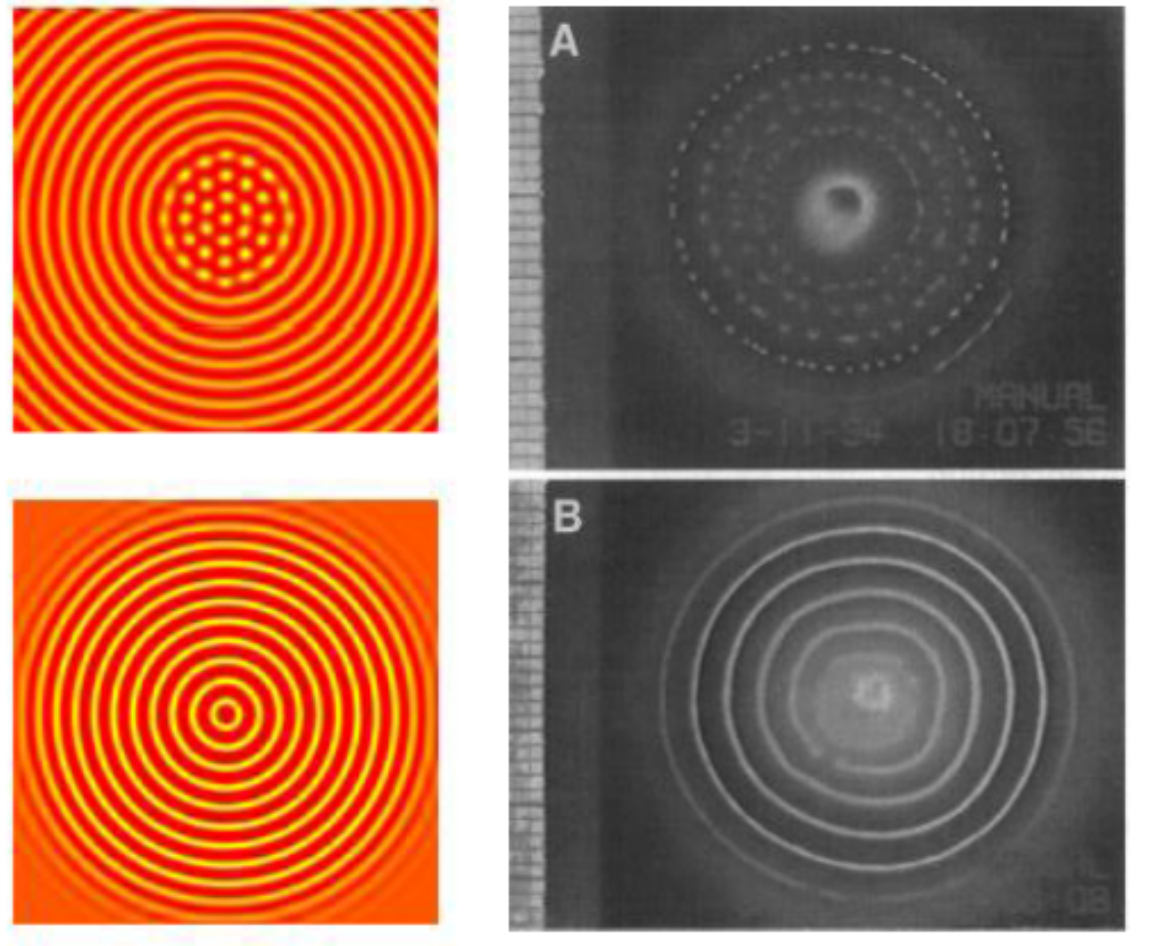}
\caption{Left frames: Simulation of the late stage behaviour of a model coupling population to phase separation, at two different sets of parameter values. (Images courtesy D. Marenduzzo; for a full description see \cite{PNAS}. Right: Experimental images of colony growth in {\em S.~Typhimurium} (From \cite{budrene}.)}
\end{center}
\end{figure}

Given the preceding discussion of the coupling between spinodal decomposition and logistic population growth, the reader should not be surprised to learn that these two ingredients alone, by creating a mechanism for microphase separation on a definite length scale, are sufficient to explain the broad phenomenology of the multi-ring and spotted patterns seen in {\em E.~coli} and {S.~typhimurium} \cite{PNAS}. (On the other hand, a separate set of patterns, seen in {\em Bacillus subtilis} and involving intricate feathery whorls, may require a quite different explanation \cite{benjacob}.) Some patterns found by solving (\ref{PNAS}) numerically are compared with the experimental ones in Figure 8. Several refinements may be necessary before the links between such a simple generic model and experimental microbiology can be fully established \cite{Brenner}. Nonetheless, the basic mechanistic picture is an appealing one: specifically it suggests that such patterns can in principle arise without chemotaxis and therefore -- if seen in future using some other organism -- they should not be taken as diagnostic of it presence.

\section{Chemotaxis without chemotactic patterns}\label{Chemo2}
A somewhat different sort of `chemotactic pattern' is often seen in bacterial colonies initiated from a point-source inoculum, under different growing conditions from the multi-ring and/or spotted patterns described above. This pattern comprises a single pronounced ring of density which slowly progresses away from the origin leaving a lower density region in its wake \cite{wolfe}. The pattern is easily understood in organisms that perform chemotaxis primarily in response to a food gradient (not a chemoattractant or repellent emitted by other individuals). As food is depleted at the centre of the colony, individuals there move outwards towards the growth front where food remains plentiful; here they can freely reproduce. This effect is generally so robust that it can be used as a rough-and-ready assay of whether a particular bacterial strain is chemotactic or not \cite{adler}. 

Notably though, in most of these assay experiments, the bacterial inoculum, which is placed on the centre of an agar gel, first penetrates the gel before the colony spreads laterally through it in a quasi-2D fashion. Moreover, in microbiological assay work the concentration of agar in the gel is not usually chosen consistently but lies anywhere within a broad window defined as `soft agar'. Systematically exploring this range, however, it was recently found experimentally that the chemotactic ring disappears if the concentration of agar in the soft gel becomes too large. For {\em E.~coli} this happens within, rather than beyond, the range of gel densities historically used in assay experiments \cite{Otti}. Hence unless care is taken to use a gel of sufficiently low density, there is a risk of false negative assay results, in which the organism does have a chemotactic phenotype, but no chemotactic pattern is seen. 

A semi-quantitative dynamical model of this effect is presented in \cite{Otti}, but the main reason behind it is already indicated by the approximate results for steady-state chemotaxis, Eqs.(\ref{beta},\ref{chemoss}) above. These equations show that the ability of an organism to develop a nonuniform population density in response to a chemical gradient depends crucially on the value of $\beta$: if this is too small, chemotaxis is ineffective. As seen from (\ref{beta}), $\beta$ depends on $\alpha_0$ and $K(t)$. The properties of the response kernel $K(t)$ are genetically determined, and unlikely to change during the course of an inoculation experiment. (This lasts hours or days, long enough for reproduction but not evolution.) So the key element in determining $\beta$ is $\alpha_0$, the tumble rate arising in the absence of any chemical gradient. In an agar gel, one expects bacteria to change orientation not only by intrinsic tumbling, but by collisions with the gel matrix. To a reasonable approximation, the latter can be represented by an increased intrinsic tumble rate $\alpha_0(C)$, which is now a function of agar concentration $C$.

It is then clear from (\ref{beta}) that the chemotactic efficiency will collapse if $\alpha_0(C)$ becomes too much larger than $\alpha_0(0)$. This is because the bilobed kernel $K(t)$ in (\ref{int}) involves a genetically fixed timescale,  $\tau_c\sim1/\alpha_0(0)$, that maximizes the chemical gradient information extractable from a straight run in the organism's normal environment. If the run length is too much reduced by gel collisions, the value of the integral is still calculated by the on-board biochemical circuit, but no longer delivers any useful information about the chemical gradient. (All contributions from the integral at time scales beyond $1/\alpha_0(C)$ become randomized by the collisional tumbles.) Although the quantitative form for $\beta$ that emerges in the high collision regime differs from (\ref{beta}) by an extra factor of $\alpha_0(0)/\alpha_0(C)$ \cite{Otti}, the result is qualitatively the same: a severe loss of chemotactic performance at high gel densities. 

Recall that in phenomenological models of chemotaxis, $\beta$ sets the value of $\chi/D$ for use when writing $V = \chi\nabla c$ in the diffusion-drift equation (\ref{pdot}). Accordingly the loss of chemotactic efficiency described above in dense porous media is not captured in any model that assumes $\chi/D$ to be independent of environmental factors. Indeed, some such models are established by analogy with isothermal gas dynamics \cite{barton,ford}. This inadvertently creates a mapping onto a detailed balance system; any possibility of circulating currents in steady state is thereby eliminated. Yet such currents can certainly be expected generically in nonuniform gels (whenever $\nabla\beta\times\nabla c \neq 0$). These concerns are relevant not only to agar gels but to many other situations where bacterial chemotaxis is performed within a porous matrix, such as sand-beds and other filtering elements in wastewater treatment plants \cite{barton,ford}: in practice such porous media are often nonuniform. Even in the case of a uniform agar gel, as detailed above and in \cite{Otti}, an implicit assumption that $\chi/D$ is independent of gel density can, by failing to account for the true statistical mechanics of bacterial motility in complex environments, give qualitatively misleading predictions.

\section{Does microbiology need statistical physics?}

Emboldened by a some interesting recent opinion pieces on the wider role of physicists in biology \cite{newman,wolgemuth2}, I will here give a personal answer to the above-stated question.

Alongside the selection of bacterial motility problems outlined above, statistical physics has recently been used to address many other questions in microbiology ranging from molecular genetics to multispecies ecology. In many of these cases, a fully stochastic description is essential. This applies especially when one is less interested in the `average' behaviour of a system than in rare events (a genetic switch being flipped \cite{geneswitch}; fixation of a gene \cite{blythe}) or in dynamic phase transitions \cite{bartek}. Indeed the importance of stochasticity is fully accepted in most branches of population biology, and links between that field and statistical physics are relatively secure and well developed \cite{blythe2}. 

On the other hand, treatments of stochasticity are rarer in models addressing chemotaxis, pattern formation in colonies, and similar areas of microbiological modelling \cite{murray}. Above we have mentioned at least one instance where neglect of noise gives qualitatively wrong physics: it fails to predict the nucleation regime of motility-induced phase separation. However there may be other such instances: omitting dynamical noise is tantamount to doing mean-field theory, and thus likely to be often unreliable, for instance whenever one is close to a continuous phase transition. 

Stochasticity apart, fundamentally different insights are offered by statistical physics and more traditional approaches to the modelling of microbiological dynamics. Statistical physics approaches usually emphasise the search for simplicity and universality of mechanism, deliberately disregarding details in the hope of finding these to be inessential. To an experimental biologist who has carefully acquired lots of data, the decision of a theoretical physicist to deliberately ignore most of it can be baffling. A more elaborate model that makes fuller use of what is known may appear preferable, even if that model also requires additional unknown parameters (whose presence might limit its falsifiability). 

Whatever modelling strategy one uses, approximations always remain necessary, and if unguided by statistical physics, these can have unintended consequences. We have met one example: in modelling chemotaxis, setting the ratio $\chi/D = \beta$ to be a fixed constant may look like a harmless simplification, but by silently imposing detailed balance, this eliminates the possibility of static circulating chemotactic currents, and all that might follow dynamically from them. It also precludes any explanation of the observed dependence of chemotactic assay behaviour on gel density \cite{Otti}. This last instance provides a clear example of how a statistical physics description is relevant to an established microbiological experiment -- the chemotactic assay -- and provides insight beyond models used previously (which assumed $\chi/D=\beta$).

A second danger, especially for detailed models with many fit parameters, is to assume that a good fit to the data offers strong evidence for the chosen model. Bayes theorem, which provides a quantitative framework for the avoidance of over-fitting \cite{mackay}, could be more widely used to guard against this danger. However, even this offers limited help in deciding between a complicated model and a simple one:  Bayesian analysis normally assumes that the data represents a true model (that one is trying to find) plus some experimental noise. As data accumulates, simple models are the first to be falsified. Yet a model that clearly does not fit the data, but nearly does so despite its gross simplifications, can provide crucial mechanistic insights. Note that this applies even if a more complicated model already exists and fits the data better.

Ultimately, choosing between complicated and simple models is a matter of individual preference, since they serve different goals. However, the understanding of generic mechanistic principles is a goal of equal importance to that of describing or even predicting experimental outcomes, and mechanistic insight can often be gleaned from the successes, and equally from the failures, of simplified models. To serve this purpose such models should at least be  consistent (both internally, and with physical law) and based on clear scientific precepts. The generation of such models, particularly where stochastic dynamics is required, is a major strength of the statistical physics approach.

Some scientists would argue that microbiology can survive purely as an experimentally driven empirical science, without quantitative theory of any kind. If this is true, then clearly microbiology does not need statistical physics.  On the other hand, if microbiology needs theory at all, as most would accept, then I believe it does need the statistical physics approach alongside the more traditional strategies that have more generally prevailed until now. 

\section*{Acknowledgements:} The author thanks Ronojoy Adhikari, Rosalind Allen, Richard Blythe, Otti Croze, Martin Evans, Davide Marenduzzo, Ignacio Pagonabarraga, Wilson Poon, Julien Tailleur and Alasdair Thompson for many discussions and collaborations pertinent to preparation of this article. He thanks Otti Croze  and Julien Tailleur for a detailed reading of the manuscript. The author is funded by a Royal Society Research Professorship and EPSRC Grant EP/EO30173.

{}
\end{document}